# Supersymmetrical bounding of asymmetric states and quantum phase transitions by anti-crossing of symmetric states

Muhammad Imran Afzal[1] and Yong Tak Lee[1,2]*

[1]Advanced Photonics Research Institute, Gwangju Institute of Science and Technology, Gwangju 500-712, Korea.
[2]School of Information and Communications, Gwangju Institute of Science and Technology, Gwangju, 500-712, Korea.
*ytlee@gist.ac.kr

**Von Neumann and Wigner theorized the bounding and anti-crossing of eigenstates. Experiments have demonstrated that owing to anti-crossing and similar radiation rates, the graphene-like resonance of inhomogeneously strained photonic eigenstates can generate a pseudomagnetic field, bandgaps and Landau levels, whereas exponential or dissimilar rates induce non-Hermicity. Here, we experimentally demonstrate higher-order supersymmetry and quantum phase transitions by resonance between similar one-dimensional lattices. The lattices consisted of inhomogeneous strain-like phases of triangular solitons. The resonance created two-dimensional, inhomogeneously deformed photonic graphene. All parent eigenstates were annihilated. Eigenstates of mildly strained solitons were annihilated at similar rates through one tail and generated Hermitian bounded eigenstates. The strongly strained solitons with positive phase defects were annihilated at exponential rates through one tail, which bounded eigenstates through non-Hermitianally generated exceptional points. However, strongly strained solitons with negative defects were effectively amplified. Supersymmetry was evident, with preservation of the shapes and relative phase differences of the parent solitons. Localizations of energies generated from annihilations of mildly and strongly strained soliton eigenstates were responsible for geometrical (Berry) and topological phase transitions, respectively. Both contributed to generating a quantum Zeno phase, whereas only strong twists generated topological (Anderson) localization. Anti-bunching-like condensation was also observed.**

The theory of supersymmetry (SUSY) describes the duality between bosons and fermions. Because our present living environment does not permit SUSY phenomena, according to the original theory, if SUSY exists, it must have broken symmetry[1-2]. Confirmation of the original SUSY is still lacking after the initial rounds of Large Hadron Collider experiments. However, SUSY techniques are helpful for identifying powerful methods for approximation and shape invariance[3]. One of the most important applications is quantum information processing or quantum simulation because of the robustness against decoherence and dissipation. It was believed that SUSY can only be observed as broken SUSY, which requires very large and complex experimental setups such as Large Hadron Collider. However, with the advent of Witten-index's non-zero value (i.e., all parent eigenstates must be annihilated)[4], unbroken SUSY may be observable in table-top quantum and dissipative systems[5], e.g., optics, semiconductors and condensed-matter systems.

Dirac cones are pure Hermitian systems that govern the intriguing properties of graphene[6], topological insulators[7] and configured condensed matter systems[8]. In optics, electronic graphene-like resonant Dirac cones[9-10] can be realized owing to the resonance of degenerate photonic states at the Brillouin zone. The resonance leads to a linear Dirac dispersion owing to the anti-crossing[11] of symmetric photonic states[12], which is clearly observable in the case of topological insulators such as bulk-state bandgaps[13]. In the case of inhomogeneous, strained waveguide arrays, the resonance leads to a pseudomagnetic field and isolated landau levels[14], where level spacing is unique to the corresponding lattices.

Non-Hermitian systems are distinguished from Hermitian systems by exhibiting Exceptional Points (EPs), which abruptly coalesce the eigenmodes without any threshold barrier[15]. The EP is a point of confusion in Hermitian systems because sudden changes can produce an imaginary potential. For example, sudden loss induces counterintuitive observations at exponential transition rates, such as lasing[16-18] and unidirectionality[19-25]. The topologies of EPs have been investigated for acoustic[26] and microwave systems[27-29]. The connections between Dirac cones and EPs and their dynamics have been theoretically investigated in photonic honeycomb lattices[30-33]. However, a recent experiment demonstrated EPs in a honeycomb. The EPs deformed the Dirac cones into lantern-like shapes[34].

Here, we experimentally demonstrate that anti-crossing in photonic graphene-like resonance leads to Hermitian and non-Hermitian bounding of eigenstates. Hermitian and non-Hermitian bounding are due to similar and non-exponential, and exponential radiation-loss rates from the resonant cones, respectively. In these transformations, the SUSY, i.e., the relative gaps of eigenstates and shapes of parent solitons, remain intact; therefore, we refer to this phenomenon as supersymmetric bounding of states. These observations have no exact experimental analogue.

We consider a soliton with phase $\varphi_i$ as mildly twisted when $\varphi_{i+1} - \varphi_i > 0$ and $\varphi_i - \varphi_{i-1} > 0$, while $|\varphi_{i+1} - \varphi_i| \neq |\varphi_i - \varphi_{i-1}|$. A soliton with phase $\varphi_i$ is treated as strongly twisted when $\varphi_{i+1} - \varphi_i < 0$ and $\varphi_i - \varphi_{i-1} > 0$ or $\varphi_{i+1} - \varphi_i > 0$ and $\varphi_i - \varphi_{i-1} < 0$. Strongly twisted solitons are consisted of positive and negative phase defects. This description is represented schematically in Fig. 1(a) and Fig. 2(a). Because of graphene-like resonance, mildly inhomogeneous strained-like modulation of phases $\varphi_i$ of the soliton lattices generate similar pseudomagnetic flux $\Phi$ owing to similar radiation rates without undergoing interband photonic transitions (evanescent tunnelling). The direction of the flux is opposite to that of the y-axis. Owing to similar radiation rate losses, all eigenstates are annihilated with similar rates in the same direction. In our experiment, the vector potential under the strain[14] can be described as $A(r) = [\varphi_i]$, and the pseudomagnetic field can be described as $B(r) = \nabla \times A(r)$. However, strong twists with positive defects generate abruptly strong fluxes, i.e., much greater than the mildly twist-induced pseudomagnetic field, owing to exponential radiation rates. The exponential radiation rates induce EPs. Thus, at EPs, the pseudomagnetic field increases exponentially. In this case, the direction of the flux or radiation is orthogonal (parallel to the x-axis) to the magnetic flux $\Phi$ of mildly twisted solitons. This type of strong or non-Hermitian twists has not been studied before. Hermitian and non-Hermitian bounds are depicted in Fig. 1 (b) and 2 (b) by vertical black and red lines. The radiated energy from mildly and strongly twisted positive defect solitons localizes as a continuum without leaving its lattice, which is responsible for quantum-phase transitions, i.e., Berry, quantum Zeno and topological phases. Two localizations of radiated energies, i.e., one part of the continuum along the x-axis owing to Hermitian and the other part along the y-axis owing to EPs, are Zeno phase, whereas localized energies along the x-axis owing to only EPs yield topological localization and effective amplification. In Fig. 1 and 2, the continuums are symbolically represented by blue filled rectangular boxes. (for details, see the text). However, we can imagine that in the presence of only mildly twisted solitons, i.e., absence of non-Hermicity, only the Berry (geometrical) phase[34-36] will be observed.

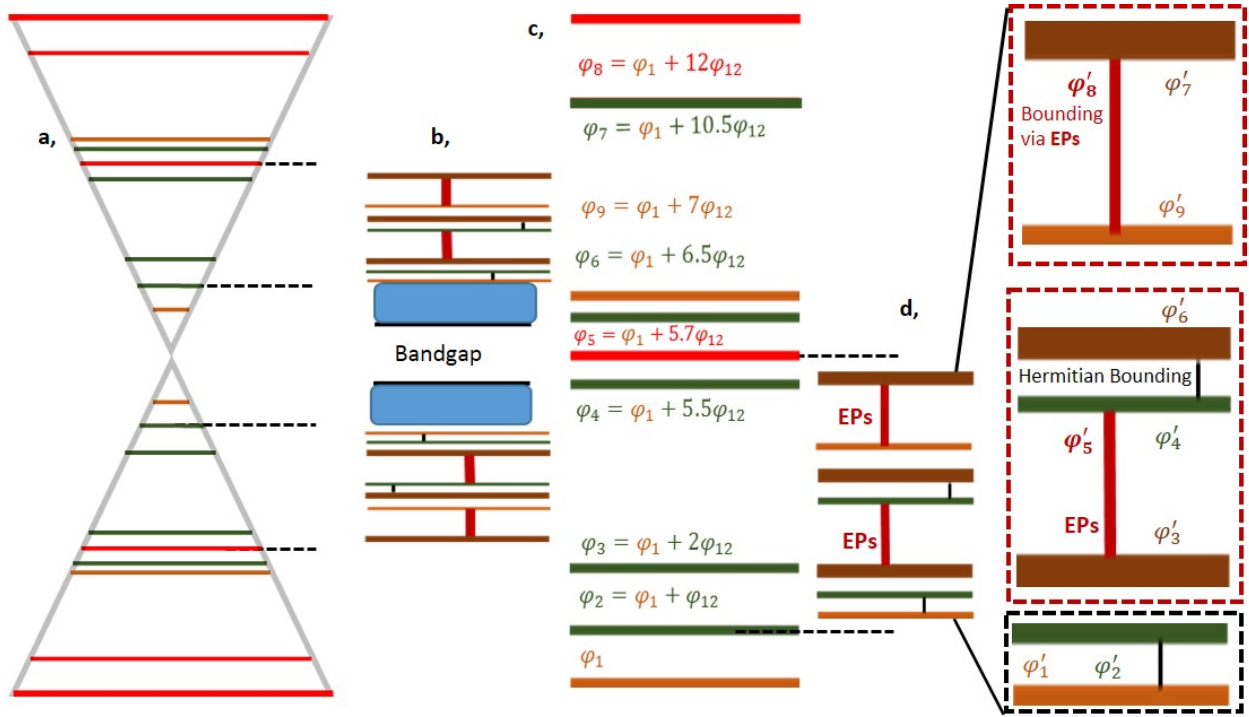

**Figure 1 | Supersymmetry, bounding of eigenstates and phase transitions through anti-crossing of nine mildly and strongly twisted parent eigenstates. a,** Graphene-like resonance of parent eigenstates. The orange, green and red lines are eigenstates of the boundary and mildly and strongly twisted positive defect soliton eigenstates, respectively. **b,** All parent eigenstates $\varphi_i$ are annihilated, and new (supersymmetric) eigenstates $\varphi_i'$ are confined between the parent regions of eigenstates $\varphi_2$ and $\varphi_5$. Owing to anti-crossing between symmetric eigenstates (two cones), bandgaps and supersymmetric bounding are generated. Mildly twisted solitons decay via radiation at similar and non-exponential rates, and are bounded Hermitianally (vertical black lines in **b** and **d**), whereas strongly twisted of positive defect solitons vanish completely by

decaying at exponential rates. Which generates EPs (vertical bold red lines) and bounded neighbouring eigenstates. Owing to EP-induced bounding, the bounded eigenstates become boundary eigenstates, indicated by dark orange lines. The energies released from the annihilations localize within the respective lattices as continuums, which are symbolically represented by rectangular boxes filled with a light blue colour. The continuums control quantum-phase transitions (see text for details). **c** and **d** are magnified presentations of **a** and **b**. In **d,** the red and black dotted rectangular boxes explicitly show the non-Hermicity- and Hermicity-induced bound states.

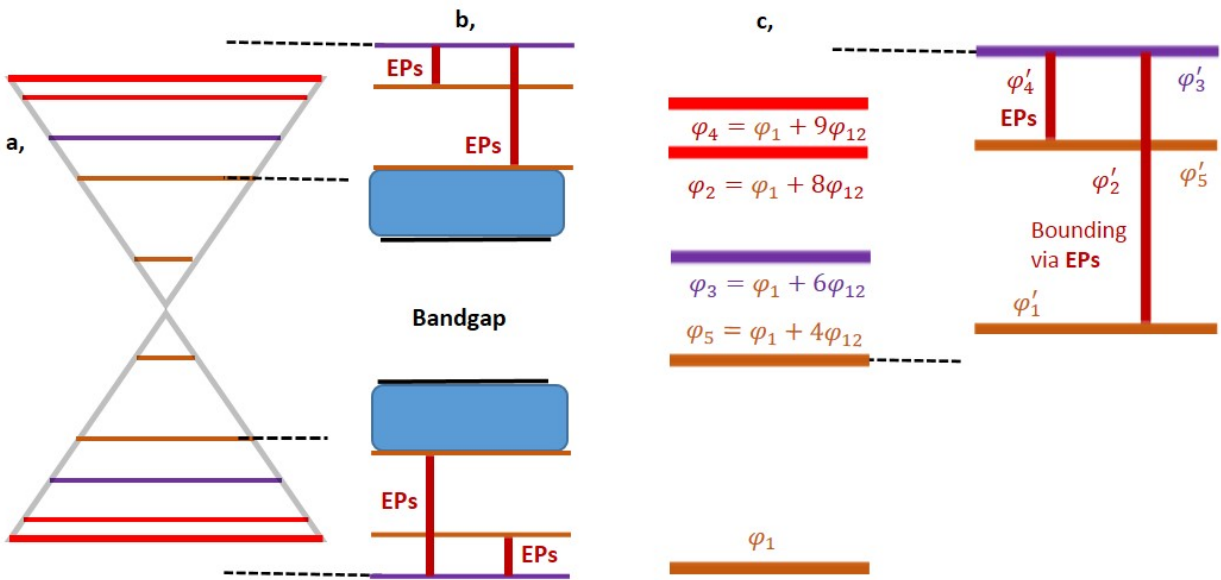

**Figure 2 | Supersymmetry, bounding of eigenstates and phase transitions through anti-crossing of five strongly twisted parent eigenstates. a,** Anti-crossing between five asymmetric parent eigenstates. The orange, purple and red lines are eigenstates of the boundary and two types of strongly twisted, i.e., negative defect and positive defect, soliton eigenstates. **b,** All parent eigenstates $\varphi_i$ are annihilated. Most of the spectrum of supersymmetric eigenstates $\varphi_i'$ is confined between the parent regions of eigenstates $\varphi_2$ and $\varphi_5$; however, the supersymmetric eigenstate $\varphi_3'$ is beyond the region of the parent spectrum. Owing to anti-crossing between symmetric eigenstates (two cones), bandgaps and supersymmetric bounding are generated in which the negative defect, strongly twisted soliton retains its parent eigenstate and therefore appears

effectively amplified. By contrast, the second type of strongly twisted solitons, i.e., positive defects, vanishes completely by decaying at exponential rates. This decay generates EPs (vertical bold red lines) and bounded neighbouring eigenstates. Owing to EP-induced bounding, the bounded eigenstates become boundary eigenstates, which are indicated by dark orange lines. The energies released from the annihilations localize within the respective lattices as a continuum, as indicated by the rectangular box filled with a light blue colour. Here, the continuum is responsible for the topological quantum phase transition. Owing to the absence of mildly twisted solitons, no berry phase can be observed (see text for details). **c** shows magnified presentations of **a** and **b**.

The experimental setup is shown in Fig. 3. The two soliton lattices occupying a wavelength range from 999.5 to 1,009.5 nm enter the 2-m loop of optical fibre. After meeting with the first coupler, two similar copies are generated owing to power splitting. The optical spectrum of the lattices detected at the output of the second coupler is captured using an optical spectrum analyser (OSA). The lattices are detected without any change in structure owing to the absence of interference/resonance at the second coupler, as shown in Fig. 4(a, c). As shown, the detected peak powers of the lattices are 3.38 and 1.15, respectively. The first lattice consists of nine solitons, where two are strongly twisted. Graphene-like resonance between the split lattices is achieved by resonances of the lattices at the second coupler by exactly matching the length of the two arms of the loop, as shown in Fig. 3.

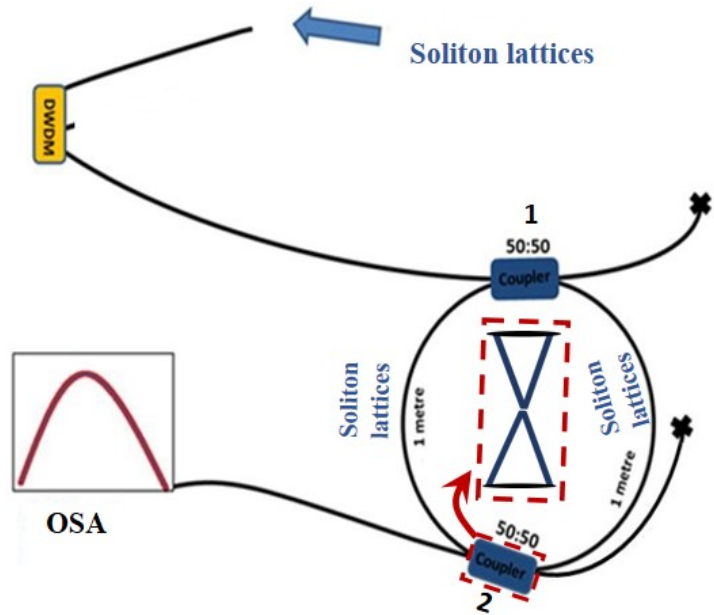

**Figure 3 | Experimental configuration.** The two soliton lattices, which have a wavelength range of 999.5 to 1,009.5 nm, enter the 2-m loop of optical fibre through the fibre dense wavelength division multiplexer (DWDM). The entire loop consists of a normal dispersion optical fibre (HI 980). The fibre exhibits zero dispersion at 1,300 nm, and the dispersion values at 980 nm are approximately -63 ps/nm/km. The fibre exhibits multimode character. A normal dispersion fibre is used to avoid modulation instability. Two copies of the input lattices with attenuated power are generated after meeting with the first 50:50 coupler. The optical spectrum of the lattices detected at the output of the second coupler is captured using an optical spectrum analyser (OSA). First, the lattices are detected without any change in structure owing to the absence of interference/resonance at the second coupler, as shown in Fig. 4(a, c). Graphene-like resonance between the split lattices is achieved by resonances of the lattices at the second coupler by exactly matching the length of the two arms of the loop, as shown in the figure. The inset is drawn to show the graphene-like resonance.

Supersymmetrically bounded eigenstates and quantum phase transitions generated from the resonance are shown in Fig. 4(b, d). The mode between the two lattices at ~1006 is unaffected by the supersymmetric transformations, indicating isolation between the two lattices. This effect also indicates that the supersymmetric transformation cannot be produced in the case of resonance of the single soliton mode or eigenstate. Therefore, the presence of more than one mode in one wave packet is required to achieve graphene-like resonance. Eigenstates of soliton lattices detected at the output of the experimental setup without resonance are shown in Fig. 4(a) and c. The first and second parent lattice comprise nine and five eigenstates ($\varphi_i$), respectively. The first lattice contains two strongly twisted solitons with positive phase defects, whereas the second lattice contains one positive and one negative phase defect soliton, as shown in Fig. 4(a) and (c), respectively. After the resonance of each of the lattices with itself, supersymmetrically bounded eigenstates are observed for the first and second lattices, as shown in Fig. 4(c) and (d), respectively, where supersymmetric eigenstates are denoted as $\varphi_i'$. Upon resonance, mildly twisted solitons radiate through one tail with similar and non-exponential radiation rates, generating Hermitian bounding, whereas a strongly twisted soliton with a positive defect radiates at exponential rate, non-Hermitianally bounding the surrounding eigenstates. However, the negative defect does not radiate, probably because of its existence between two positive defects.

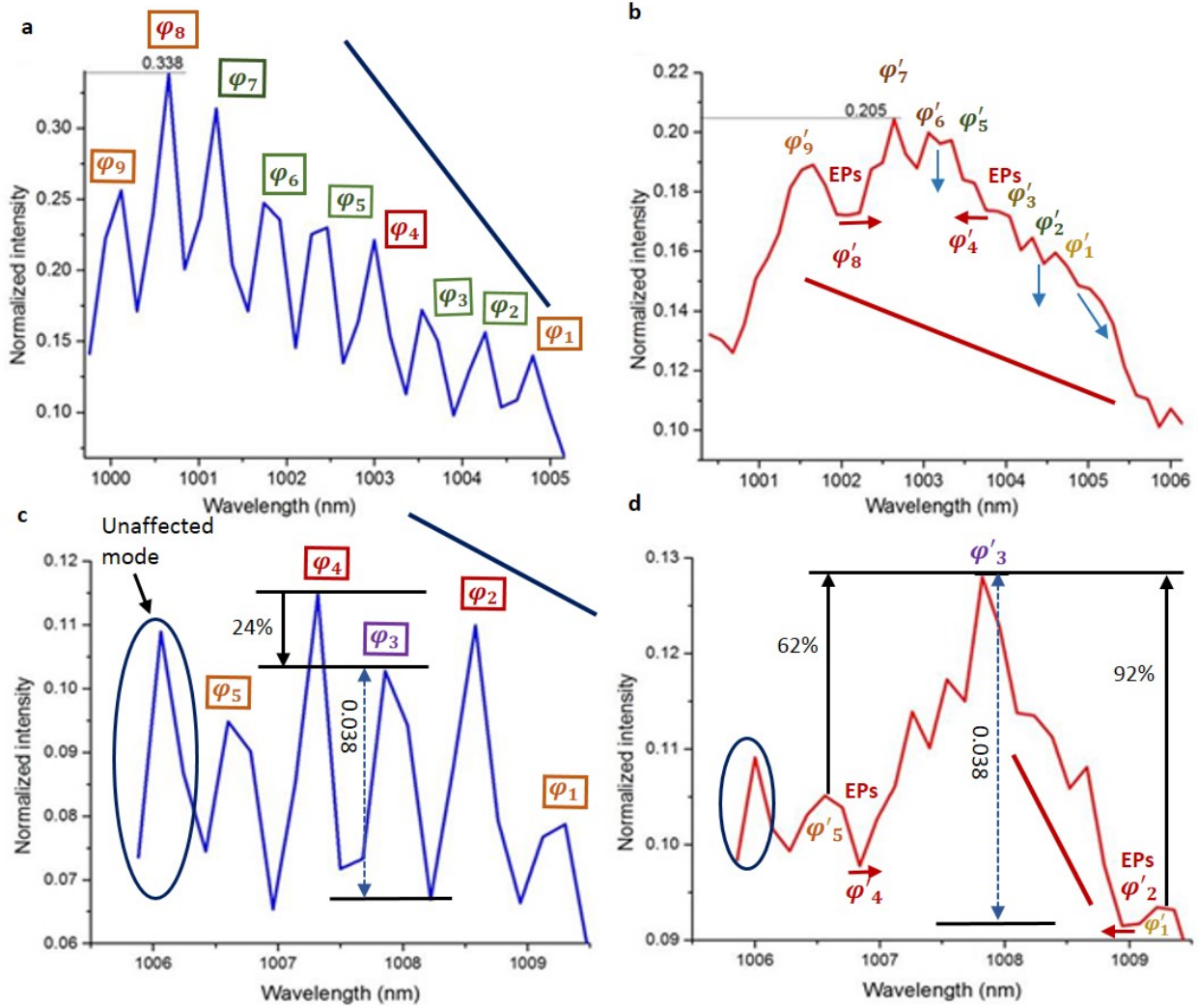

**Figure 4 | Eigenstates of soliton lattices detected at the output of the experimental setup before and after generation of supersymmetric bounding of eigenstates and effective amplification. a, c,** The first and second parent lattice comprise nine and five eigenstates ($\varphi_i$), respectively (for details, see Fig. 1 and 2). **c,** The parent lattice (**c**) does not contain a Hermitian soliton but does contain two types of strong twists, i.e., both positive and negative defects. **b,** Supersymmetric eigenstates ($\varphi'_i$) and their Hermitian and non-Hermitian bounding. The thin blue arrows indicate the direction of radiation loss of Hermitian solitons. Thick red arrows indicate directions of the radiation owing to non-Hermicity. All parent eigenstates of Hermitian solitons are annihilated and generate supersymmetric eigenstates ($\varphi'_i$), which are bounded Hermitianally. However, strongly twisted, positive defect solitons $\varphi_4$ and $\varphi_8$ are annihilated at exponential rates and vanish completely,

whereas their positions remain localized, i.e., $\varphi'_4$ and $\varphi'_8$. The non-Hermitian annihilation generates EPs, which bound both sides of the eigenstates. The peak power is reduced to 2.04 µW. Instead of escaping, owing to the pseudomagnetic field, the energies released by the annihilations localize to form a continuum and are responsible for quantum-phase transitions. The continuum decreases the slope of the parent spectrum from 1.31 to 0.43. The new slope is highlighted by the straight red line. This decrease in the slope of the redistributed eigenspectrum is interpreted as the quantum Zeno phase. The topological phase features are due to EPs, whereas the geometrical phase features are due to Hermicity. **d,** Only non-Hermitian but very different dynamics are observed for the negative defect soliton ($\varphi'_3$), which sustains its parent energy at 0.038 while surrounded by positive defect-induced EPs on both sides. Thus, the defect with phase $\varphi'_3$ appears effectively amplified. However, the spectrum of the eigenstates overall is topologically localized owing to localization of the continuum along the y-axis, as evident by the increase in slope from 0.54 to 1.91; these slopes are represented by blue (**c**) and red straight lines, respectively. A mode represented as an 'unaffected mode', located at a wavelength of ~1006 nm in **c** and **d**, is unaffected by the transformations, which indicates the isolation of the first and second lattices from each other.

Mainly, Hermitian soliton's radiated energy accumulates along the x-axis to form a flat-band-like continuum[14,34,37], whereas non-Hermitian soliton's energy accumulates along the y-axis. The former is responsible for the geometrical phase transition, whereas the latter is responsible for the topological phase transition. A combination of geometrical and topological phase transitions enables the observation of the quantum Zeno phase, as shown in Fig. 4(b). However, there is no doubt that following our idea quantum Zeno effect can also be observed in all Hermitian solitonic system. The non-Hermitian annihilations generate EPs, which bound both sides of the eigenstates. The peak power is reduced to 2.04 µW. Here, the localizations of one part of the continuum along the x-axis are due to the Hermitian, whereas the localizations of the other part the along the y-axis (at the bottom of $\varphi'_5$, $\varphi'_6$ and $\varphi'_7$) owing to EPs are revealed as the quantum Zeno phase. The collective effect is represented by the decrease in slope from 1.31 to 0.43, as indicated by the blue and red straight lines in Fig. 4(a) and (b), respectively. The presence of a negative defect between positive defects enables the observation of the topological localization of the whole spectrum. Here, the localization of energies at the bottom of $\varphi_3'$ is revealed as topological localization; the parent and final slopes are represented by blue and red straight lines in Fig. 4(c) and (d), respectively. In contrast to the positive defect, instead of radiating, the negative defect was effectively amplified. The slope of the parent spectrum increased from 0.54 to 1.91. This observation allows us to introduce an interesting postulate: if a SUSY potential cannot preserve the eigenspectrum of the parent system in a specific dimension, then it will localize the eigenspectrum of the parent potential to another dimension to obey the SUSY. This aspect is also important for understanding the increase and decrease in (effective) dimensions in any local system or evolution (growth) of new dimensions in a system, including biology. In both Fig. 4(b) and (d), the parent shapes of the solitons and their phase (spacing) difference are relatively preserved, except for positive defects, i.e., the parent eigenstates $\varphi_i$ are relatively preserved as $\varphi_i'$. Therefore, the bounded states are considered as supersymmetrically bounded states.

Our observed topological localization is different from the conventional Anderson localization in optics; however, the shapes are in agreement with Anderson localization[38] and the anti-Zeno effect[39-40] in condensed matter and effective amplification[41] in quantum optics. Effective amplification is a necessary element for the experimental implementation of quantum computation, quantum detection and Schrodinger's cat-based amplification of the quantum states. A comparison of Figs. 4(c) and (d) in terms of the effective amplification of modes instead of the redistribution of the energies of the parent eigenstates demonstrates that the mode at 1,007.86 nm is amplified from 0.103 to 0.128, and the effective amplification ranges from -24% (as indicated by the black downward arrow) to 62% and 92% (as indicated by the black upward arrows). The relative power (0.038) of the signal at 1,007.86 nm is nearly identical for the parent and SUSY transformed system.

The redistribution of energies in the localized continuums allowed us to observe the reduced intensity fluctuations in the original SUSY systems, as shown in Fig. 4(b) and (d), as well as the ascending power distribution of continuums, as shown in Fig. 5, which is analogous to anti-bunched clustering. In particular, the redistributed energies in the quantum Zeno phase profile (distribution of red circles in Fig. 5(a)) clearly show antibunched photonic bands or clustering of the states. A theoretical study of supersymmetrically generated photonic bands in disordered optical potentials (waveguide arrays) without anti-crossing was recently reported[42]. However, in principle, the results reported here are different.

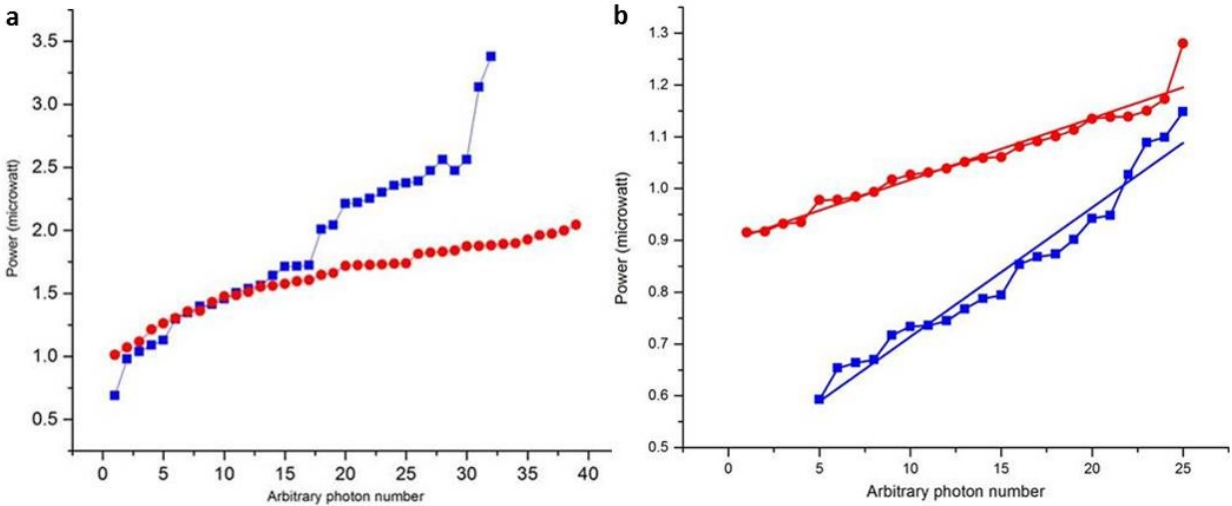

**Figure 5 | Comparison of the ascending power distribution of continuums of parent soliton lattices, Fig. 4(a) and Fig. 4(c), with the continuums of SUSY bounded lattices, Figs. 4(b) and 4(b). a,** The power for an arbitrary number of photons in the first parent lattice is shown in blue, whereas the power redistribution in the SUSY lattice is shown in **red**. The red graph shows the formation of three clusters. Interestingly, Fig. 4(b) also shows three bounds. **b,** The power for an arbitrary number of photons in the second parent lattice is shown in blue, whereas the power redistribution in the SUSY lattice is shown in red. Fig. 4(d) shows two boundings; however, owing to a negative defect, even the presence of two boundings results in the formation of a single lattice-like structure. The similarity in behaviour between Fig. 4 and 5 can be understood by considering the localization of the energies in the continuum, as represented by the lines with slopes in Fig. 4(b, d).

**Conclusion and summary**

We observed supersymmetrically bounded eigenstates through Hermicity and non-Hermicity in soliton lattices with weakly and strongly twisted solitons or inhomogeneous eigenstates, achieved by the anti-crossing of asymmetric states. The non-Hermicity or abrupt annihilations of a few modes in a lattice of a very large number of Hermitian modes may play a crucial role in quantum phase transitions, whereas the Hermitian modes are only responsible for the Berry phase. The preservation of the parent shapes and phase difference of the solitons are clear evidence of SUSY, whereas the large number of annihilations of the parent eigenstates is evidence of higher-order transformations. Different topologies of the EPs are also observed: some are matched with the spawning of EPs in photonic graphene[34], whereas others are matched with EPs in acoustics[26] and microwave billiards[27-29]. We believe that the results will prove very helpful for envisioning strongly bounded eigenstates via non-Hermicity beyond the Von Neumann and Wigner's Hermitianally bounded states. Our results may also be visualized as Hermitian[14] and non-Hermitian bounding of Landau levels. The results may also be important for further study of many-body quantum computation or simulation and its limitations using anti-crossing[43]. For example, Hermitian part in quantum Zeno phase transition may helpful for understanding many body-quantum information processing. While our observed topological Anderson localization somehow also supports the idea that Anderson localization cause failure of adiabatic quantum computation[44].

**References**

1. Martin, S. P. A Supersymmetry primer' in: `Perspectives in Supersymmetry. G.L. Kane (Ed.), World Scientific, Singapore, and arXiv:9709356 [hep-ph].

2. Dine, M. Supersymmetry and String Theory: Beyond the Standard Model (Cambridge University Press, 2007).

3. Cooper, F., Khare, A. & Sukhatme, U. Supersymmetry and quantum mechanics. *Phys. Rep.* **251**, 267 (1995).

4. Witten, E. Constraints on Supersymmetry Breaking. *Nucl. Phys.* **B202,** 253-316 (1982).

5. Tomka, M., Pletyukhov, M., & Gritsev, V. Supersymmetry in quantum optics and in spin-orbit coupled systems. *Sci. Rep.* **5**, 13097 (2015).

6. Castro Neto, A. H., Guinea, F., Peres, N. M. R., Novoselov, K. S. & Geim, A. K. The electronic properties of graphene. *Rev. Mod. Phys.* **81**, 109–162 (2009)

7. Hasan, M. Z. & Kane, C. L. Colloquium: topological insulators. *Rev. Mod. Phys.* **82**, 3045 (2010)

8. Tarruell, L., Greif, D., Uehlinger, T., Jotzu, G. & Esslinger, T. Creating, moving and merging dirac points with a fermi gas in a tunable honeycomb lattice. *Nature* **483**, 302–305 (2012).

9. Kane, C. L. & Mele, E. J. Size, Shape, and Low Energy Electronic Structure of Carbon Nanotubes. *Phys. Rev. Lett.* **78**, 1932–1935 (1997).

10. Zhang, Y., Tan, Y.-W., Stormer, H. L. & Kim, P. Experimental observation of the quantum Hall effect and Berry's phase in graphene. *Nature* **438**, 201–204 (2005).

11. Von Neumann, J. & Wigner, E. Über merkwürdige diskrete Eigenwerte. *Phys. Z.* **30**, 465–467 (1929).

12. Sakoda, K. Proof of the universality of mode symmetries in creating photonic dirac cones. *Opt. Express* **20**, 25181–25194 (2012).

13. Rechtsman, M. C. et al. Photonic Floquet topological insulators. *Nature* **496**, 196–200 (2013).

14. Rechtsman, M. C. et al. Strain-induced pseudomagnetic field and photonic landau levels in dielectric structures. *Nature Photon.* **7**, 153–158 (2013).

15. Heiss, W. The physics of exceptional points. *J. Phys. A* **45**, 444016 (2012).

16. Liertzer, M. et al. Pump-induced exceptional points in lasers. *Phys. Rev. Lett.* **108**, 173901 (2012).

17. Brandstetter, M. et al. Reversing the pump-dependence of a laser at an exceptional point. *Nature Comm.* **5**, 4034 (2014).

18. Peng, B. et al. Loss-induced suppression and revival of lasing. *Science* **346**, 328–332 (2014).

19. Guo, A. et al. Observation of PT-symmetry breaking in complex optical potentials. *Phys. Rev. Lett.* **103**, 093902 (2009).

20. Makris, K., El-Ganainy, R., Christodoulides, D. & Musslimani, Z. H. Beam dynamics in PT symmetric optical lattices. *Phys. Rev. Lett.* **100**, 103904 (2008).

21. Longhi, S. Bloch oscillations in complex crystals with PT symmetry. *Phys. Rev. Lett.* **103**, 123601 (2009).

22. Lin, Z. et al. Unidirectional invisibility induced by PT-symmetric periodic structures. *Phys. Rev. Lett.* **106**, 213901 (2011).

23. Feng, L. et al. Experimental demonstration of a unidirectional reflectionless parity-time metamaterial at optical frequencies. *Nature materials* **12**, 108–113 (2013).

24. Peng, B. et al. Parity-time-symmetric whispering-gallery microcavities. *Nature Physics* **10**, 394–398 (2014).

25. Chang, L. et al. Parity-time symmetry and variable optical isolation in active-passive-coupled microresonators. *Nature Photon.* **8**, 524–529 (2014)

26. Ding, K. et al. The emergence, coalescence and topological properties of multiple exceptional points and their experimental realization. arXiv:1509.06886 (2015).

27. Dembowski, C. et al. Experimental Observation of the Topological Structure of Exceptional Points. *Phys. Rev. Lett.* **86**, 787 (2001).

28. Dembowski, C. et al. Observation of a Chiral State in a Microwave Cavity. *Phys. Rev. Lett.* **90**, 034101 (2003).

29. Lee, S-Y. et al. Geometric phase around multiple exceptional points. *Phys. Rev. A* **85**, 064103 (2012).

30. Szameit, A. P T-symmetry in honeycomb photonic lattices. *Phys. Rev. A* **84**, 021806 (2011).

31. Ramezani, H. Kottos, T. Kovanis, V. & Christodoulides, D. N. Exceptional-point dynamics in photonic honeycomb lattices with PT symmetry. *Phys. Rev. A* **85**, 013818 (2012).

32. Yannopapas, V. Spontaneous PT-symmetry breaking in complex frequency band structures. *Phys. Rev. A* **89**, 013808 (2014).

33. Monticone, F. & Alu, A. Embedded photonic eigenvalues in 3d nanostructures. *Phys. Rev. Lett.* **112**, 213903 (2014).

34. Zhen, B., Hsu, C. W., Igarashi, Y., Lu, L., Kaminer, I., Pick, A., Chua, S.-L., Joannopoulos, J. D. & Soljacic, M. Spawning rings of exceptional points out of Dirac cones. *Nature* **525**, 354 (2015).

35. Chan, C., Hang, Z. H. & Huang, X. Dirac dispersion in two-dimensional photonic crystals. *Adv. Optoelectron.* **2012**, 313984 (2012).

36. Mei, J., Wu, Y., Chan, C. & Zhang, Z.-Q. First-principles study of Dirac and Dirac-like cones in phononic and photonic crystals. *Phys. Rev. B* **86**, 035141 (2012).

37. Huang, X., Lai, Y., Hang, Z. H., Zheng, H. & Chan, C. Dirac cones induced by accidental degeneracy in photonic crystals and zero-refractive-index materials. *Nature materials* **10**, 582–586 (2011).

38. Roati, G. et al. Anderson localization of a non-interacting Bose-Einstein condensate. *Nature* **453**, 895–898 (2008).

39. Fischer, M. C., Gutiérrez-Medina, B. & Raizen, M. G. Observation of the quantum Zeno and anti-Zeno effects in an unstable system. *Phys. Rev. Lett.* **87**, 040402-040404 (2001),

40. Fujii, K. & Yamamoto, K. Anti-Zeno effect for quantum transport in disordered system. *Phys. Rev. A* **82**, 042109 (2010).

41. Zavatta, A., Fiuráˇsek, J. & Bellini, M. A high-fidelity noiseless amplifier for quantum light states. *Nature Photon.* **5**, 52–60 (2010).

42. Yu, S., Piao, X., Hong, J. & Park, N. Bloch-like waves in random-walk potentials based on supersymmetry. *Nature Comm.* **6**, 8269 (2015).

43. Li, W., Tao, T., Bo, G., Cheng, Z. & Guang-Can, G. Experimental realization of non-adiabatic universal quantum gates using geometric Landau-Zener-Stückelberg interferometry. *Sci. Rep.* **6**, 19048 (2016).

44. Altshuler, B., Krovi, H. & Roland, J. Anderson localization makes adiabatic quantum optimization fail. *Proc. Natl Acad. Sci. USA* 107, 12446–12450 (2010).

**Acknowledgements**

This work was partially supported by a "Systems biology infrastructure establishment grant" and the "Ultrashort Quantum Beam Facility Program" through a grant provided by the Gwangju Institute of Science and Technology in 2015.

**Author Contributions**

M.I. conceived the idea, designed and performed the experiments and analysed the results. Y.T. provided insightful advice and encouraged M.I. during the investigation and supervised the project. M.I. and Y.T. wrote the manuscript. All authors reviewed the manuscript.

**Competing Interests Statement**

The authors declare that they have no competing financial interests.